\documentstyle[cltp]{article}
 
\def\Journal#1#2#3#4{{#1} {\bf #2}, #3 (#4)}
 
\def\NIM{\em Nucl. Instrum. Methods}

\def\etal{{\it et al.}}

\def\mco{\multicolumn}

\def\be{\begin{equation}}
\def\ee{\end{equation}}
\def\bea{\begin{eqnarray}}
\def\eea{\end{eqnarray}}
 
\input ./epsf.tex

\begin{document}
 
\title{The {\v C}erenkov Correlated Timing Detector: \\
     Beam Test Results from Quartz and Acrylic Bars }
 
\author{ H. Kichimi$^\star$, Y. Sugaya, H. Yamaguchi and Y. Yoshimura }
 
\address{KEK, Tsukuba, Japan }
 
\author{S. Kanda, S. Olsen, K. Ueno, G. Varner}
 
\address{ University of Hawaii, Honolulu, Hawaii, USA}
 
\author{T. Bergfeld, J. Bialek, J. Lorenc, M. Palmer, G. Rudnick, M. Selen  }
 
\address{Department of Physics, University of Illinois 
at Urbana-Champaign, Urbana, Illinois, USA }
 
\author{ T. Auran, V. Boyer, K. Honscheid }
 
\address{ The Ohio State University, Columbus, Ohio, USA}
 
\author{N. Tamura, K. Yoshimura }
 
\address{Okayama University, Okayama, Japan }
 
\author{C. Lu, D. Marlow, C. Mindas, E. Prebys }
 
\address{Princeton University, Princeton, New Jersey, USA }
 
\author{M. Asai, A. Kimura, S. Hayashi }
 
\address{Hiroshima Institute of Technology, Hiroshima, Japan }

\twocolumn[\maketitle\par

\abstract{ Several prototypes of a {\v C}erenkov Correlated Timing (CCT) Detector 
have been tested at the KEK-PS test beam line.  
We describe the results for {\v C}erenkov light yields and
timing characteristics from quartz and acrylic bar prototypes. 
A {\v C}erenkov angle resolution is found to be 15 mrad 
at a propagation distance of 100 cm with a 2 cm thick quartz bar prototype. }

]
 
\vspace*{1ex} 

\section{Introduction}

The {\v C}erenkov Correlated Timing (CCT) Detector is a new particle identification concept based on
precision timing measurements to determine the {\v C}erenkov angle of photons emitted
by particles passing through a transparent radiator~\cite{mats-cct}.
In order to investigate the detector concept, several prototypes have been tested 
at the KEK-PS test beam line, as tabulated in Table 1.
Material tests have been done in parallel with measurements of the optical 
characteristics~\cite{mats-rich}.
The beam test data are well reproduced by Monte Carlo simulation,
so that the feasibility of the CCT detector concept has been
demonstrated.
We describe the results mainly from a quartz bar prototype 
and summarize the results from other prototypes for comparison. 
      
\section{Experiment}

\subsection{Beam test setup}\label{subsec:setup}
\begin{figure}
\centerline{ \epsfxsize=7.5cm \epsfbox{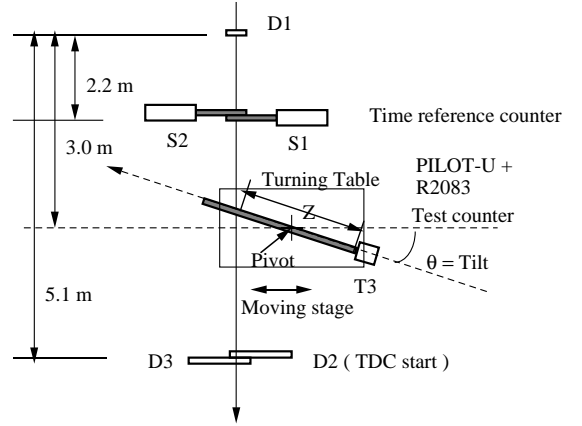} }
\caption{Beam test setup }
\label{fig:setup}
\end{figure}

The beam test setup at the KEK-PS T1 beam line is shown in Figure~\ref{fig:setup}.
CCT prototypes are placed on a rotating table which is mounted on a moving stage, so that
the beam position and the crossing angle can be adjusted independently.
The accuracies in positioning the prototype counter were 0.5 cm along the counter 
axis and 1$^o$ in tilt angle, respectively.
The beam was defined by two sets of beam definition counters D1 and D2$\times $D3, which were 
separated by about 5 m. Their areas were 2$\times $2 cm$^2$ and 1$\times $1 cm$^2$,
respectively, thus the divergence of the beam was expected to be about 2 mrad. 
{\v C}erenkov light yields and timing characteristics were investigated as function of  
beam position (Z) and tilt angle ($\theta$) of the counter, using a 1.5 GeV/c $\pi^-$ beam.

\subsection{CCT prototypes}\label{subsec:Zygo CCT pmt}
The tested CCT prototypes are tabulated in Table~\ref{tab:CCTpro}.
Details of the material characteristics are described in Ref.~\cite{mats-rich}.  
The Acrylic bars had been annealed for 10 hours at 80$^o$C to release internal stress
after machining and polishing. 
A support for the Zygo quartz bar CCT was carefully designed 
to avoid a sizable loss of photons during total internal reflection. 
As the edges were machined very sharply and fragile,
the mechanical contacts were made of Teflon and designed to 
touch only the surfaces and not the edges with the contact areas minimized.
This point is very important to realize an attenuation length of several 10's of m,
whose reflectivity per bounce is expected to be on the order of 0.9995. 
One end of the counter was viewed by a phototube and the opposite end
was blackened to avoid reflection.
 
\begin{table}\begin{center}\caption{CCT prototypes}\label{tab:CCTpro}
\vspace{0.4cm}
\begin{tabular}{|c|c|c|c|c|} \hline
 Prot. & (1) & (2) & (3) & (4) \\ \hline
 Test Run & Mar.95 & \mco{3}{|c|}{ Nov.94 } \\ \hline
 Prep. & KEK   & U.I.   & \mco{2}{|c|}{ KEK } \\ \hline
 Mater.& Quartz & Quartz  & Acrylic & Acrylic  \\ 
 Prod. & Zygo & S. Finish & Mitsubishi & Kuraray \\ \hline
 Index & 1.47 &1.47 &1.51 &1.51  \\ \hline
 Thickness & 2 cm & 4 cm  & 3 cm & 5 cm \\
 Width & 4 cm & 4 cm & 4 cm & 6 cm \\ 
 Length & 120 cm & 100 cm & 150 cm & 100 cm \\ \hline
 Couplant & Viscasil & \mco{3}{|c|}{ OKEN 6262A } \\ \hline
 PMT  & \mco{4}{|c|}{ H2431 } \\ \hline
 $N_o$(pe/cm) & 136 & 120 & 66 & 82 \\  
 $\lambda_{att}$(m) & 13.6 & 1.8 & 1.43 & 1.45 \\ 
 $\sigma_{CCT}$(ps) & 80 & 80 & 105 & 95 \\ 
 $\delta_{\Theta}$(mrad) & 15 & 15 & 20 & 20 \\ \hline
\end{tabular}
\end{center}
\end{table}
 
 The same photomultiplier tube (Hamamatsu H2431) was used to measure the light yield
 and timing characteristics of all of the prototypes. 
 The rise time and the transit time spread ($\sigma_{CCT}$) 
 are 0.7 ns and 160 ps.
 The gain of the tube was calibrated by using a laser pulser (Hamamatsu PLP-02) in single photon mode. 
 A gain of 3.65 ADC counts per photoelectron was obtained at an operating voltage of 3100V,
 with an error of 10$\%$. 
 The H2431 is 2" tube, with an effective photocathode diameter of 46 mm,
 and a borosilicate window with a bi-alkali photocathode.
 It is sensitive in the wavelength range between 300 nm to 500 nm 
 and has a peak quantum efficiency of 25\% at about 350 nm. 
 The photocathode sensitivity curve $QE(\lambda)$ is taken into 
 account in MC simulation. 

\subsection{ {\v C}erenkov light yield of Zygo bar CCT }
\label{subsec:CCT yield}

\begin{figure}
\centerline{ \epsfxsize=6.5cm \epsfbox{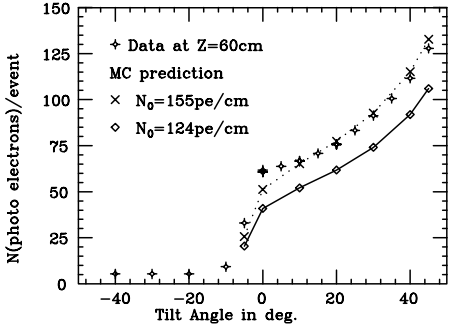} }
\caption{ {\v C}erenkov light yield vs. $\theta$ }
\label{fig:Zygo N p.e.}
\end{figure}

The signals from the CCT counters were read out by CAMAC TDC and ADC modules, 
with a least count of 25 ps and 0.25 pC, respectively. 
Light yields were evaluated from the ADC data and the gain described above.
Figure~\ref{fig:Zygo N p.e.} shows the light yield measured for a Zygo quartz bar CCT,
as a function of tilt angle $\theta$ at Z= 60 cm.
The light yield shows a rapid decrease from $0^o$ to
a cutoff around $\theta$= -7$^o$ corresponding to a critical angle. However, 
there is seen a non zero light yield below $\theta$= -10$^o$, 
where none of {\v C}erenkov photons could reach the readout phototube. 
These are due to {\v C}erenkov photons from $\delta$-rays produced by the beam passage.
The details of $\delta$-ray {\v C}erenkov contribution will be discussed in Section~\ref{sec:delta-ray}.

The dotted line shows a Monte Carlo prediction with a 
{\v C}erenkov quality factor $N_o$ of 155 pe/cm, obtained by a fit to the data,
as described below and in Section~\ref{section:MC}.
Our measurement of $N_o$ is 20\% larger than the results measured 
at SLAC~\cite{BaBar} (124 pe/cm) and Princeton Univ~\cite{PU} (121 pe/cm) with
cosmic rays. The solid line shows the prediction with $N_o$= 124 pe/cm.
This excess is attributable to the photocathode sensivity
and/or some contribution from $\delta$-ray {\v C}erenkov photons from the 1.5GeV $\pi ^- $ beam.

$N_0$ was obtained from the data 
using Equation~\ref{eq:it1},
\begin{equation}
N_{pe}^{meas} = \frac{N_0 sin^2{\Theta_c}d} { cos\theta} \! ~G
\label{eq:it1}
\end{equation}
where $\Theta_c$ is {\v C}erenkov angle, d is thickness of CCT
counter, and $\theta$ is a tilt angle of the counter to the beam. 
G is a collection efficiency of photons 
in a given CCT counter geometry, and 
it is evaluated by a Monte Carlo simulation using Equation~\ref{eq:it2},
\begin{equation}
G = \frac{N_{pe}(MC)} {\int\!\frac{dN}{d\lambda}\!~QE(\lambda)~d\lambda}\! 
               \frac{cos\theta}{d}\! 
\label{eq:it2}
\end{equation}
$N_{pe}(MC)$ is number of photoelectrons expected by MC simulation,
$\frac{dN}{d\lambda}\!$ is {\v C}erenkov photon density and $QE(\lambda)$ is photocathode
sensitivity. 

\begin{figure}
\centerline{ \epsfxsize=6.5cm \epsfbox{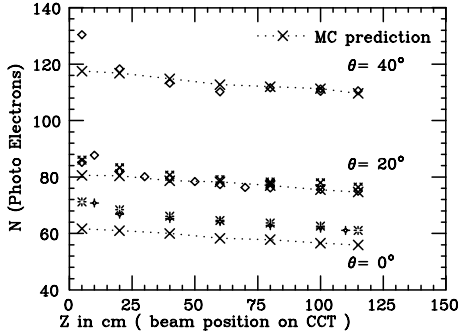} }
\caption{ {\v C}erenkov Attenuation in Zygo bar }
\label{fig:Zygo att}
\end{figure}

Figure~\ref{fig:Zygo att} shows {\v C}erenkov light yield as a function of distance Z
at three tilt angles: $\theta$= 0$^o$, 20$^o$ and 40$^o$.
Reflectivity per bounce $\epsilon_{ref}$ is found to be 0.9996$\pm$0.0003 by a fit.
The dotted lines show the predictions with $N_0$=155 pe/cm and $\epsilon_{ref}$= 0.9996. 
The peaks at shortest distances are not predicted by this MC simulation. 
The attenuation lengths in the region of Z$\ge$20 cm  were found to be 10.7$\pm$0.4 m, 
13.6$\pm$2.2 m and 43.8$\pm$26.7 m, respectively.

\subsection{ Timing characteristics of the Zygo bar CCT }\label{subsec:Zygo CCT}

 The discriminator level was set at 50 mV, which corresponded to
 a few photoelectron signal. 
 The start time jitter for each run was evaluated to be 40$\sim$50 ps, and it was subtracted 
 in quadrature to obtain the intrinsic time resolution for each of the CCT prototypes.

\begin{figure}
\centerline{ \epsfxsize=6.5cm \epsfbox{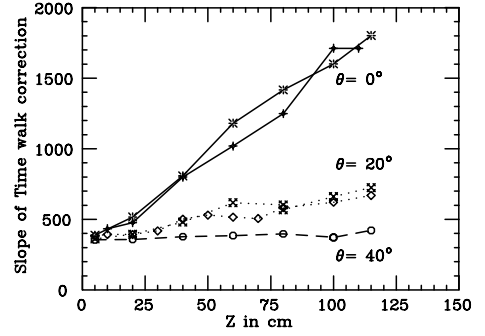} }
\caption{ Slope parameters of time walk as a function of Z }
\label{fig:CCT twc}
\end{figure}

\begin{figure}
\centerline{ \epsfxsize=6.5cm \epsfbox{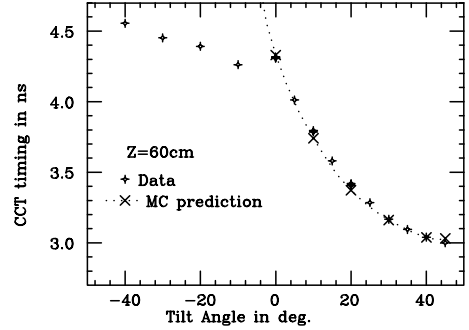} }
\caption{ CCT timing vs. $\theta$ at z=60cm }
\label{fig:CCT timing}
\end{figure}

 The start time for each event was evaluated using a pair of start scintillation counters S1 and S2.
  Each TDC measurement was corrected from time walk by using a formula $T_C \ =\ (TDC - TDC_{start}) + a - b/\sqrt{ADC} $.
  Figure~\ref{fig:CCT twc} shows the fitted slope parameter $b$ as function of
  Z and $\theta$.
  The parameter b shows very little dependence on Z at $\theta$= 40$^o$,
  while it shows a strong correlation with Z at $\theta$= 20$^o$ and 0$^o$.
  This is due to contamination from $\delta$-ray {\v C}erenkov photons.
  The resultant time resolution was investigated as a function of $b$,
  and a single value was chosen to correct all of the TDC data.
  It corresponded to a value of about 400 found at $\theta$= 40$^o$ over all Z, to which 
  all of the values converge at Z$\le$15 cm, as shown in Figure~\ref{fig:CCT twc}. 
  Figure~\ref{fig:CCT timing} shows corrected timing $T_C$ as a function of $\theta$ at Z= 60 cm. 
  Prediction by Monte Carlo simulation shows good agreement with the data, with an {\it rms} of 20 ps.
  In the region of $\theta$ $\le$ -10$^o$, $T_C$ is almost independent of $\theta$, 
  suggesting a uniform cosine distribution due to $\delta$-ray {\v C}erenkov photons.  
  
\begin{figure}
\centerline{ \epsfxsize=6.5cm \epsfbox{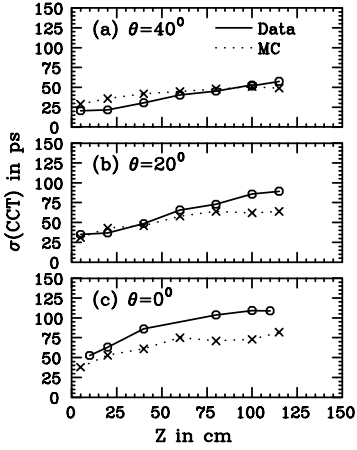} }
\caption{ $\sigma_{CCT}$ vs. Z at (a) $\theta$= 40$^o$, (b) 20$^o$ 
       and (c) 0$^o$, compared with Monte Carlo prediction }
\label{fig:CCT sigma}
\end{figure}

  Figure~\ref{fig:CCT sigma} shows time resolution $\sigma_{CCT}$ as a function of Z
  at the three tilt angles (a) 40$^o$, (b) 20$^o$ and (c) 0$^o$, respectively.
  The $\sigma_{CCT}$'s being 30 ps at Z= 5$\sim$10 cm, 
  increasing with Z up to 50 ps, 80 ps and 105 ps at Z=100 cm, respectively.
  The MC predictions ( dotted lines ) show good agreement with the data.
  A systematic discrepancy is observed at $\theta$= 0$^o$, which increases 
  up to $\sim$20\% at Z= 100 cm.
 
\section{ Monte Carlo Simulation of CCT prototype }\label{section:MC}

  According to the given functions~\cite{mats-cct} for the {\v C}erenkov photon density 
  and for the dispersion,
  {\v C}erenkov photons are emitted along the path of particle passage through the CCT counter. 
  Each photon propagates by total internal reflection, at a critical angle 
  given by its wavelength. 
  After many bounces on the counter surface and going through the couplant,
  some fraction of the {\v C}erenkov photons arrive at read out phototube
  and emit photoelectrons. 
  The data of M.Griot in Ref.~\cite{BaBar} was used to simulate photon absorption in the quartz bar.
  In this simulation, only two parameters were assumed to be free: reflectivity per bounce 
  ($\epsilon_{ref}$), and a normalization factor ($Q$). 
  This $Q$ corresponds to a peak quantum efficiency (nominally $\sim$25\%). 
  These parameters were found to be 0.9996$\pm$0.0003 for $\epsilon_{ref}$,
  and 25\% ($N_o$=155 pe/cm) for $Q$ by fitting to the data.  
  
  We take into account the timing characteristics of
  the phototube as described in section~\ref{subsec:Zygo CCT pmt}.
  In addition, we introduce a pulse height fluctuation for single photoelectron signals~\cite{mats-cct}~\cite{pmt}.
  For this, a Gaussian distribution was assumed, having an average
  of 3.65 ADC counts and a standard deviation of a half of the average.
  To simulate the time walk effect, a discriminator was also included.
  An assumption for $\sigma_{TTS}$ of 160 ps is well reproduced for single photoelectron signals, 
  by applying the time walk correction to Monte Carlo data.

\section{ Effects from $\delta$-rays }\label{sec:delta-ray}
  
 A Monte Carlo simulation using GEANT code has been carried out to study the effects of
 $\delta$-rays on quartz bar CCT counters, with 1.5GeV $\pi ^- $ beam.
 The fraction of the photons from $\delta$-rays is expected to be 15\% at normal beam incidence.
 The number of beam {\v C}erenkov photons shows a Gaussian (Poisson) distribution,
 while that from $\delta$-ray {\v C}erenkov photons shows a rapidly decaying tail peaked at zero. 
 The observed ADC distribution is a sum of those two components.
 
 The angular divergence of the photons from $\delta$-rays is about 100$^o$ in FWHM, 
 showing an almost uniform cosine distribution with respect to the beam direction.
 The spread in arrival times of photons is expected to be proportional to their propagation length.  
 In contrast, those from a beam particle are well collimated
 with a divergence of less than 1$^o$ in FWHM. 
 The two components behave very differently as function of Z and $\theta$, so that
 they are expected to separate at longer Z and at smaller $\theta$.
 
\begin{figure}
\centerline{ \epsfxsize=8.5cm \epsfbox{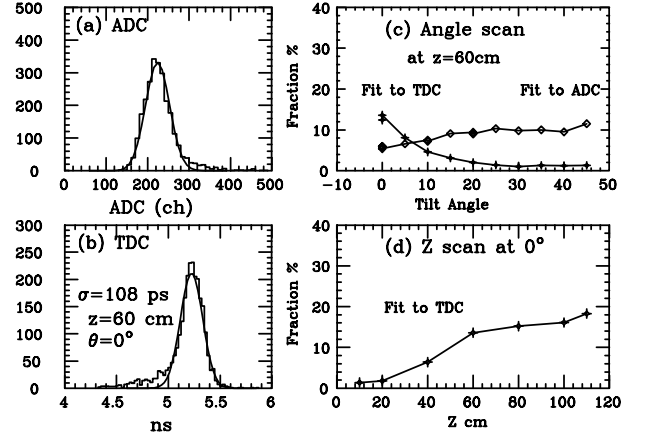} }
\caption{ (a) ADC and (b) TDC distributions at z=60cm and at $\theta$=0$^o$,
 and fraction of $\delta$-ray events 
 as function of (c) $\theta$ at Z= 60 cm 
 and (d) Z at $theta$= 0$^o$, estimated by a Gaussiun 
 fit to ADC and TDC distributions. } 
\label{fig:delta-ray}
\end{figure}

 Figure~\ref{fig:delta-ray}(a) shows ADC data at Z=60 cm and $\theta$=$0^o$.
 The tail in the ADC and the TDC distributions were separated by a fit to a Gaussian plus a tail.
 The results are shown in Figure~\ref{fig:delta-ray}(c).
 10$\sim$12\% of the photons are attributable to $\delta$-rays.
 Subtracting this contribution, a quality factor $N_o$ is calculated to be 136 pe/cm. 

 Figure~\ref{fig:delta-ray}(b) shows a TDC distribution after time walk correction.
 The tail at earlier times corresponds to $\delta$-ray {\v C}erenkov photons.
  Figure~\ref{fig:delta-ray}(d) shows the results
 as a function of Z at $\theta$=$0^o$.
 This indicates that 16\% of events are triggered by $\delta$-ray 
 {\v C}erenkov photons at Z= 100 cm. 
 As seen in Figure~\ref{fig:delta-ray}(b), about 10\% of events would be triggered at a time 
 earlier by 2$\sigma$ than expected  from beam {\v C}erenkov photons.
 Such a wrong trigger probability due to $\delta$-ray {\v C}erenkov photons may depend
 on discrimination level, as well as Z and $\theta$.

\section{ Results from other prototypes }

In Table~\ref{tab:CCTpro} are listed
the measured {\v C}herenkov quality factors, attenuation length and time
resolution at $\theta$= 20$^o$, and {\v C}erenkov angle resolution ($\delta_{\Theta_C}$)
at $\theta$= 0$^o$.
The attenuation length of Surface Finishes (SF) quartz bar is measured to be 1.8 m, which is mainly
attributable to its beveled edges. The acrylic bars show a short attenuation length
of about 1.4 m and quality factors of 66$\sim$82 pe/cm.

\begin{figure}
\centerline{ \epsfxsize=6.5cm \epsfbox{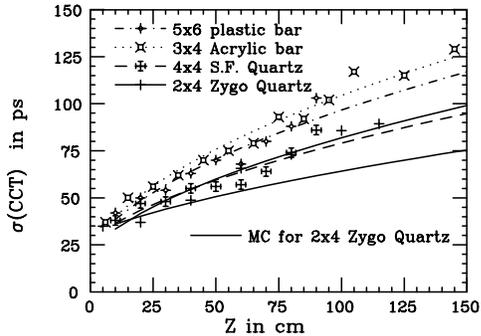} }
\caption{ $\sigma_{CCT}$ vs. Z for quartz and acrylic CCT 
          at $\theta$=20$^o$}
\label{fig:CCT_sigma}
\end{figure}

Figure~\ref{fig:CCT_sigma} shows the measured time resolution of a Zygo bar at $\theta$=20$^o$,
as a function of Z, together with other prototype counters.
These are separated into groups of quartz and acrylic bars. The curves are fits to a function
$\sigma_{CCT}=\sqrt{a+b\times L}$, assuming that the resolution is a function of 
the number of bounces of the photons on the surface.
The measured resolutions of the quartz bars are worse than the MC prediction by about 25\%.
As described in section~\ref{subsec:CCT yield},~\ref{subsec:Zygo CCT} and in Ref~\cite{mats-rich},
the optical characteristics of the Zygo bar are extremely good,
while the obtained time resolutions are almost the same as for the SF bar. 
The acrylic bars show fairly good time resolutions,
in spite of a poorer surface quality and a cutoff at lower wavelength.
These are worse only by about 20\% as shown in Figure~\ref{fig:CCT_sigma}. 
The fraction of $\delta$-ray {\v C}erenkov photons  
are also found to be about 16\% for the acrylic bars. 

\begin{figure}
\centerline{ \epsfxsize=6.5cm \epsfbox{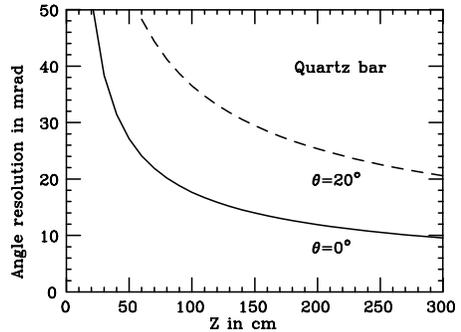} }
\caption{ {\v C}erenkov angle resolution vs. Z }
\label{fig:Angle separation}
\end{figure}

Figure~\ref{fig:Angle separation} shows {\v C}erenkov angle resolution as a function of
Z at $\theta$= $0^o$ and $20^o$, using the fitted curves to the Zygo quartz bar data.
A {\v C}erenkov angle resolution of 15 mrad is expected at a propagation distance of 100 cm and at
normal incidence, which corresponds to 3$\sigma$-$\pi/K$ separation up to a momentum of 1.4 GeV/c.   

\section{Summary}
  The {\v C}erenkov light yield and timing characteristics have been investigated
  for quartz and acrylic bar prototypes with a 1.5 GeV/c $\pi^-$ beam, 
  and the feasibility of the CCT detector concept 
  has been demonstrated.
   
 The results for a Zygo quartz bar CCT are in good agreement with the previous measurements with cosmic rays.
 A {\v C}erenkov quality factor $N_o$ of 136 pe/cm and a reflectivity per bounce of 0.9996$\pm$0.0003
 have been obtained. 
 The measured {\v C}erenkov light yield and CCT timing performance are well reproduced by Monte Carlo
 simulation, with the exception of $\sigma_{CCT}$ at smaller tilt angles and at larger distances.
 A Surface Finishes quartz bar showed almost the same $\sigma_{CCT}$ as the Zygo bar,
 in spite of its shorter attenuation length.
 
 Acrylic bars have been investigated and found to show a better than expected $\sigma_{CCT}$,
 worse only by about 20\% than the quartz bars,
 in spite of a smaller $N_o$ of 66$\sim$82 pe/cm, a shorter attenuation 
 length of 1.4 m and a marginal surface quality.
 The number of bounces of photons on the surface seems to be a key factor in determining 
 the time resolution. However, requirements  for the surface quality 
 seem not to be strict for CCT performance.

We gratefully acknowledge the support of the Department of Energy, the
National Science Foundation, and the A. P. Sloan Foundation.
We would also like to express our thanks to Prof. S.Iwata and
Prof. F.Takasaki at KEK for their support of this work.


\end{document}